\documentclass[cameraready]{NetSec2014}
\usepackage[hyphenbreaks]{breakurl}
\usepackage{hyperref}
\begin{document}


\title{Analysis of Docker Security}

\author{Thanh Bui\\
        Aalto University School of Science \\
	\texttt{thanh.bui@aalto.fi}}
\maketitle


\begin{abstract}
Over the last few years, the use of virtualization technologies has increased dramatically.  This makes the demand for efficient and secure virtualization solutions become more obvious. Container-based virtualization and hypervisor-based virtualization are two main types of virtualization technologies that have emerged to the market. Of these two classes, container-based virtualization is able to provide a more lightweight and efficient virtual environment, but not without security concerns. In this paper, we analyze the security level of Docker, a well-known representative of container-based approaches. The analysis considers two areas: (1) the internal security of Docker, and (2) how Docker interacts with the security features of the Linux kernel, such as SELinux and AppArmor, in order to harden the host system. Furthermore, the paper also discusses and identifies what could be done when using Docker to increase its level of security.

\vspace{3mm}
\noindent KEYWORDS: Containers, Docker, Security

\end{abstract}


\section{Introduction}
The last decade has seen an explosion of development in the area of virtualization technologies, which allow the partitioning of a computer system into multiple isolated virtual environments. The technologies offer substantial benefits that have been driving their development rapidly. One of the most common reasons for adopting virtualization technologies is \textit{server virtualization} in data centers. With server virtualization, an administrator can create one or more virtual system instances on a single server. These virtual systems operate as real physical servers and can be rented out on a subscription basis. Amazon EC2, Rackspace, and DreamHost are some popular instances of such data center service providers. Another common use is for \textit{desktop virtualization}, where one computer can run several OS instances. Desktop virtualization provides support for applications that can run only on a specific OS.

The growth in the use of virtualization technologies promotes the demand for a virtualization solution which can provide dense, scalable, and secure user environments. A large number of virtualization solutions have emerged to the market. They can be classified into two major classes: container-based virtualization and hypervisor-based virtualization. Of these two classes, container-based virtualization is able to provide a more lightweight and efficient virtual environment. It allows ten times more virtual environments to run on a physical server compared to hypervisor-based virtualization \cite{burniske_containers:_????}. However, container-based virtualization also comes with security concerns. 

In this paper, we analyze the security level of Docker \cite{_what_????}, a well-known representative of container-based virtualization approach. We consider two areas: (1) the internal security of Docker, and (2) how Docker interacts with the security features of the Linux kernel, such as SELinux and AppArmor, in order to harden the host system. The analysis examined the internal security of Docker based on the level of isolation Docker can provide to its virtual environments. The interaction between Docker and the security features of the kernel was estimated based on how the features are supported by Docker. To the best of our knowledge, Docker is a relatively new technology, and this is one of the first analyses of this kind that focus on its security aspects. 

The paper is structured as follows: Section~\ref{sec:virtualization_approaches} provides a high-level view of the two classes of virtualization solutions. Section~\ref{sec:docker_overview} gives an overview of Docker and its underlying technologies. Section~\ref{sec:docker_security} presents our analysis of Docker security, and then in Section~\ref{sec:discuss}, we discuss the security level of Docker and what could be done to raise its level of security. The paper concludes with a summary in Section~\ref{sec:conclusion}.


\section{Virtualization Approaches}
\label{sec:virtualization_approaches}
Most of the virtualization technologies can be classified into two major approaches: container-based virtualization and hypervisor-based virtualization. The former provides virtualization at the operating system level, while the latter provides virtualization at the hardware level. Each of the approaches has its own advantages and disadvantages, which are described in this section. 

\begin{figure}[t]
	\begin{center}
		\scalebox{.4}{\includegraphics{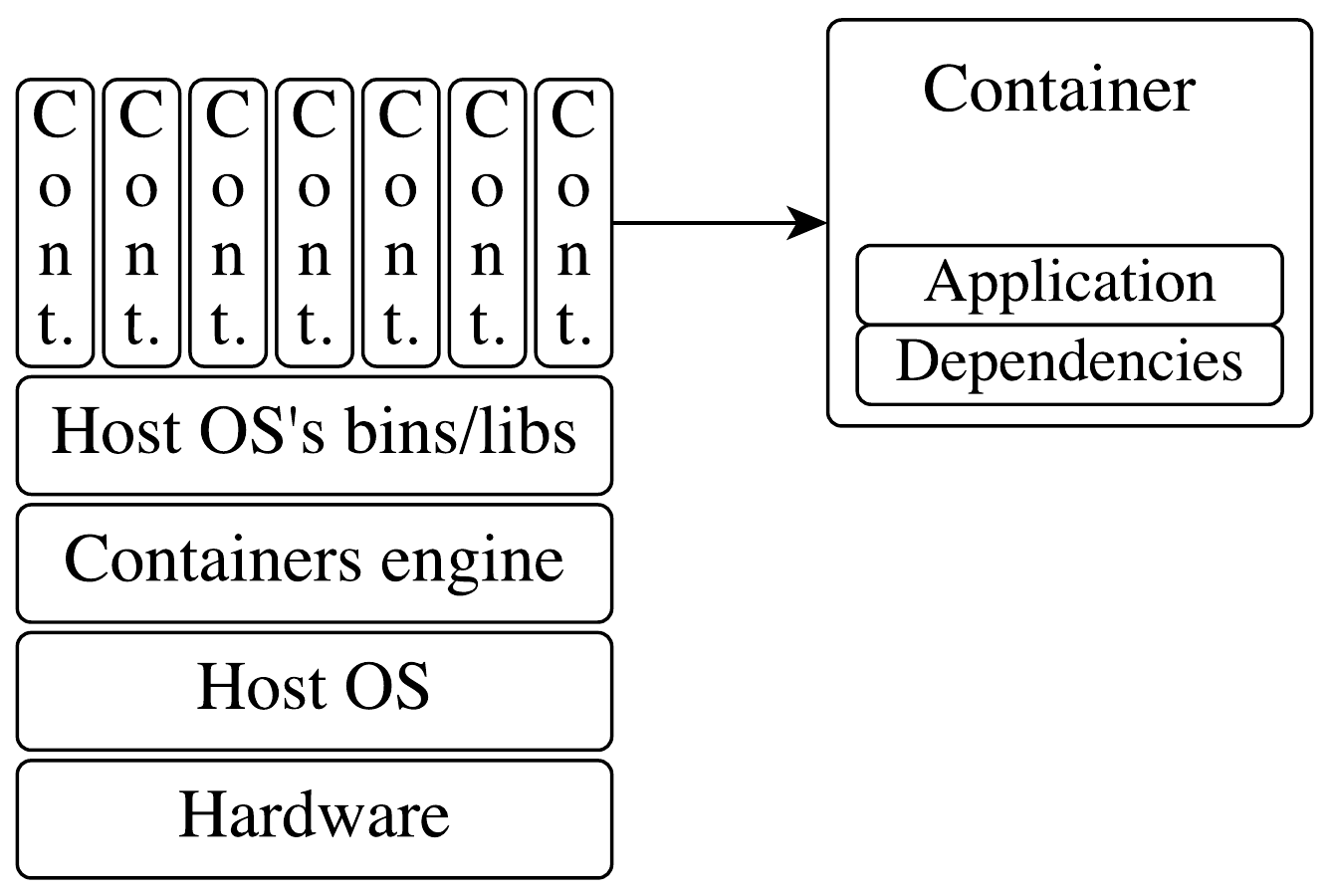}}
    		\caption{Architecture of Container-based Virtualization}
    		\label{fig:containers_architecture}
  	\end{center}
\end{figure}

\textbf{Container-based virtualization} is a lightweight virtualization approach using the host kernel to run multiple virtual environments. These virtual environments are often referred to as \textit{containers}. Linux-VServer \cite{soltesz_container-based_2007}, OpenVZ \cite{_openvz_????}, and Linux Container (LXC) \cite{_lxc_????} are the three main representatives of this approach. The general architecture of a container-based virtualization solution is illustrated in Fig.~\ref{fig:containers_architecture}. Container-based virtualization virtualizes at the operating system level, thus allowing multiple applications to operate without redundantly running other operating system kernels on the host. Its containers look like normal processes from outside, which run on top of the kernel shared with the host machine. They provide isolated environments with necessary resources to execute applications. These resources can be either shared with the host or installed separately inside the container.

\begin{figure}[t]
	\begin{center}
		\scalebox{.4}{\includegraphics{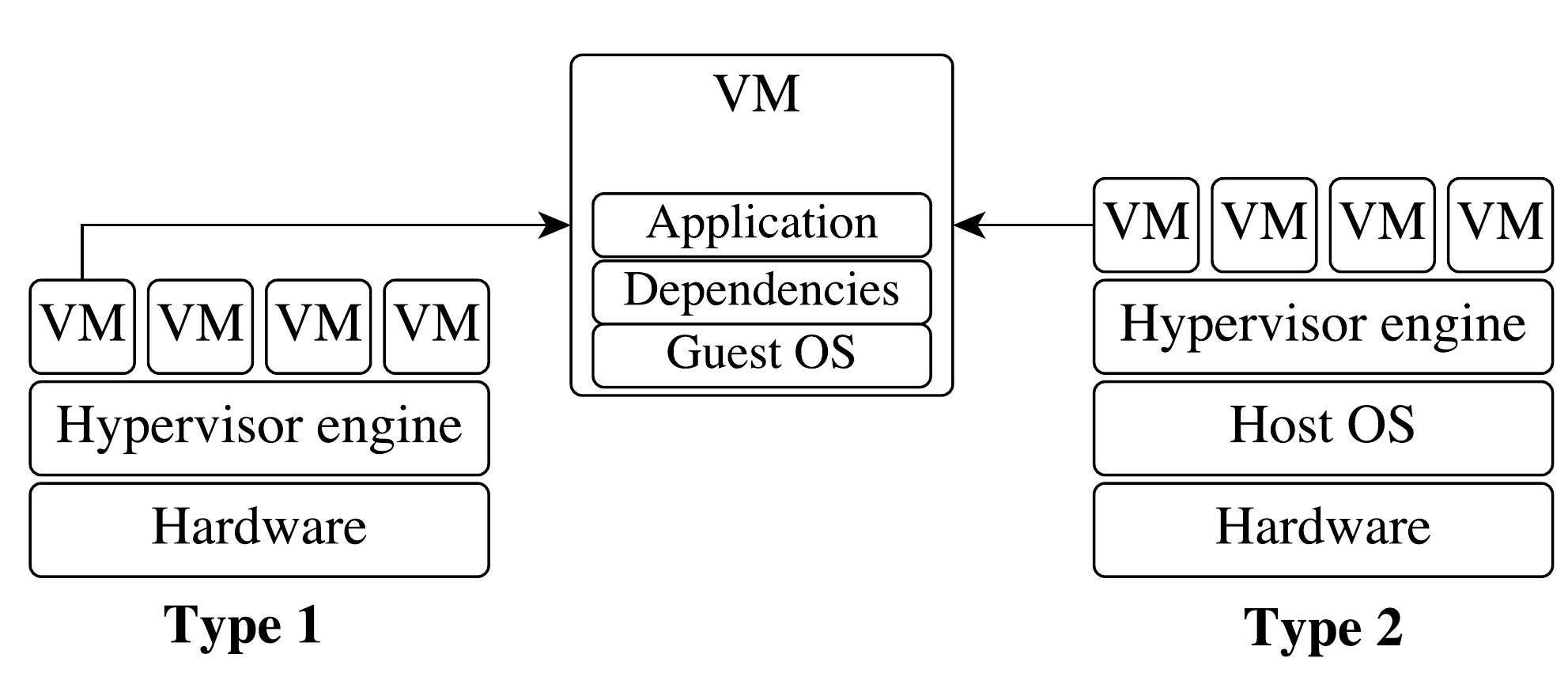}}
		\caption{Architecture of Hypervisor-based Virtualization}
		\label{fig:hypervisor_architecture}
	\end{center}
\end{figure}

\textbf{Hypervisor-based virtualization} solutions provide virtualization at the hardware level. In contrast to container-based virtualization, a hypervisor establishes complete virtual machines (VMs) on top of the host operating system (Fig.~\ref{fig:hypervisor_architecture}). Each virtual machine comprises of not only an application and its dependencies, but also an entire guest OS along with a separate kernel. There are two classes of hypervisors: the \textit{Type 1} hypervisor, also known as the bare metal hypervisor, which works directly on top of the underlying hardware of the host, and the \textit{Type 2} hypervisor, also known as the hosted hypervisor, which works on top of the host operating system \cite{merkel_docker:_2014}. Xen \cite{anderson_xen_2009} is an example of the former, while KVM \cite{kivity_kvm:_2007} is of the latter. Since the \textit{Type 1} hypervisor does not include an extra layer of the host OS, it provides better performance than the \textit{Type 2} hypervisor.

The differences in the architecture bring some benefits to container-based virtualization over hypervisor-based virtualization. First, container-based virtualization can provide \textit{higherer density} of virtual environments. Since a container does not include an entire OS, the size and the required resources to run an application in a container are less than that of a VM running the same application. As a result, more containers than traditional virtual machines can be deployed on the same host. Secondly, container-based virtualization also offers \textit{better performance}. This has been demonstrated by experiments in some studies \cite{xavier_performance_2013, regola_recommendations_2010, padala_performance_2007, felter_updated_2014}. These studies show that the performance of container-based virtualization is better than with hypervisor-based virtualization in most cases, and it is almost as good as native applications. 

However, despite all of the mentioned advantages, container-based virtualization is unable to support a variety of environments in the way hypervisor-based virtualization does since all the environments of the containers must be of the same type as that of the host. For example, Windows containers cannot be run on top of a Linux host.

\section{Docker Overview}
\label{sec:docker_overview}
Docker is an open source container technology with the ability "to build, ship, and run distributed applications" \cite{_what_????}. It has been used in some popular applications, such as Spotify, Yelp, and Ebay. 

Although container technologies have been around for more than a decade, Docker - a relatively new candidate - is currently one of the most successful technologies since it comes with new abilities that earlier technologies did not possess. First, it provides interfaces to simply and safely create and control containers. Secondly, developers can pack applications into lightweight Docker containers which can operate on almost anywhere without modification. Furthermore, Docker can deploy more virtual environments than other technologies can on the same hardware \cite{burniske_containers:_????}. Last but not least, Docker cooperates well with third-party tools, which simplify the management and deployment process of Docker containers. DevOps tools, such as Puppet \cite{_puppet_????}, Ansible \cite{_ansible_????}, and Vagrant \cite{_vagrant_????} can integrate with Docker, thus making Docker containers to be easily deployed to a cloud. Moreover, many orchestration tools, such as Mesos \cite{hindman_mesos:_2011}, Shipyard \cite{_shipyard_????}, and Kubernetes \cite{_kubernetes_????}, also support Docker containers. These tools provide an abstract layer of resources management and scheduling over Docker.

Docker consists of two major components: \textit{Docker engine} and \textit{Docker Hub}. The former is an open source virtualization solution, while the latter is a Software-as-a-Service platform for sharing Docker images. The following sections describe in details these two components.

\subsection{Docker Engine}
\begin{figure}[t]
	\begin{center}
		\scalebox{.4}{\includegraphics{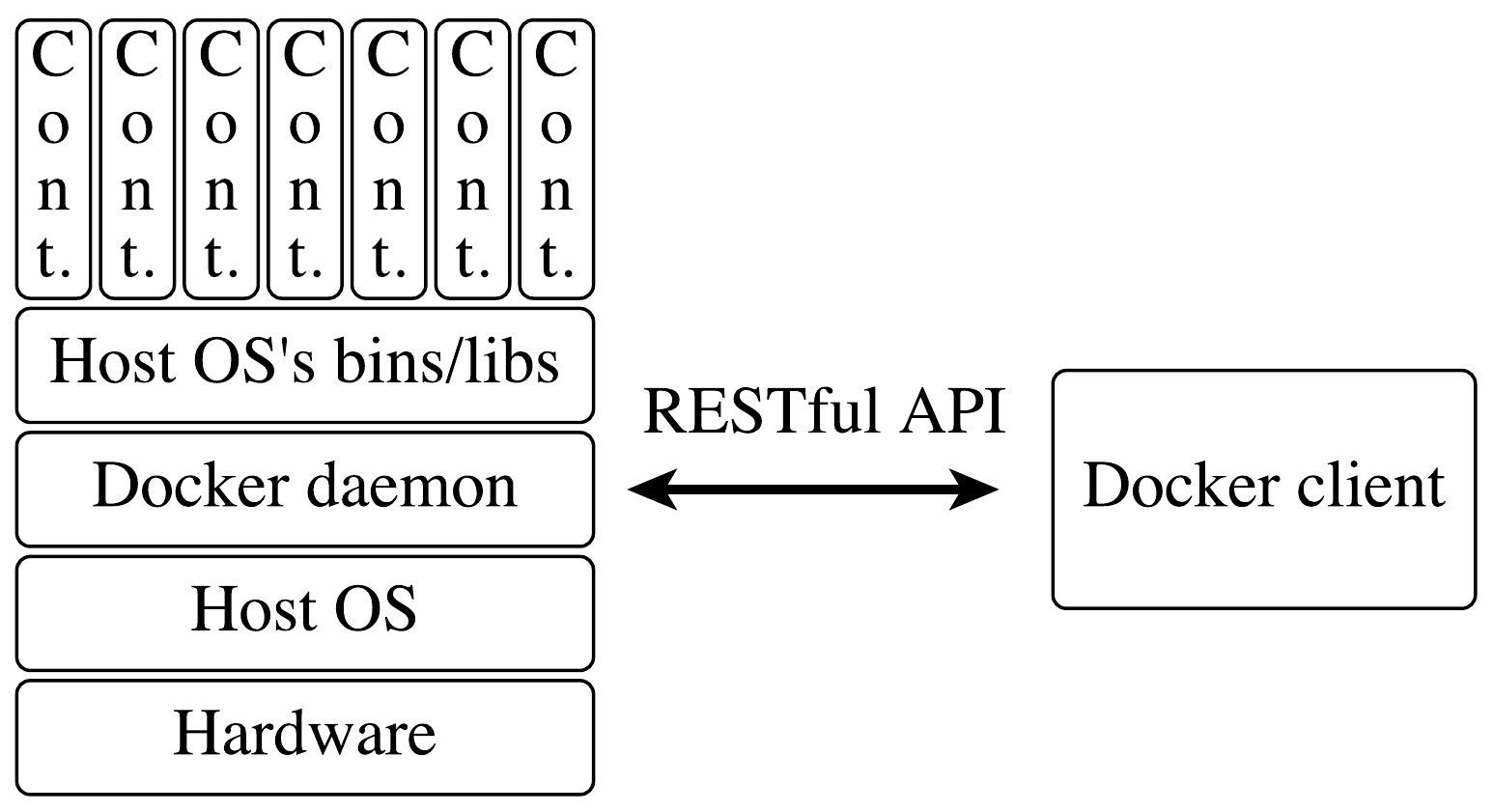}}
		\caption{Architecture of Docker engine}
		\label{fig:docker_architecture}
	\end{center}
\end{figure}

Docker engine is a lightweight and portable packaging tool \cite{_what_????} which relies on container-based virtualization. Therefore, the architecture of the Docker engine (Fig.~\ref{fig:docker_architecture}) is similar to that of container-based virtualization in general. The Docker containers run on top of the \textit{Docker daemon} which is in charge of executing and managing all of the Docker containers. The \textit{Docker client}, which provides an user interface for interacting with containers to Docker users, accepts commands from the users and then sends it to the Docker daemon through RESTful APIs. Using this method of communication enables the Docker client to run on the same host as the containers, or even on different hosts.

\subsubsection*{Docker Container}
Docker used to commoditize LXC to create Docker containers. Since version 0.9, Docker has replaced LXC with libcontainer \cite{_libcontainer_????} - their own virtualization format - as the default container environment since Docker community desires not to depend on a third-party package. However, with either LXC or libcontainer, namespaces, cgroups, union file system, and Docker images are still the major underlying technologies to implement Docker containers.  

Docker takes advantages of two Linux features, \textit{namespaces} and \textit{cgroups}, to safely create virtual environment for its containers. The cgroups, or control groups, provide mechanism for accounting and limiting the resources which the processes in each container can access. The namespaces wrap the operating system resources into different instances. The use of these instances gives the processes running inside a container the illusion that they have their own resources. Currently, Docker uses five namespaces to provide each container with a private view of the underlying host system \cite{j._walsh_are_????}: mount, hostname, inter-process communication (IPC),  process identifiers (PID), and network. Each of them works on specific types of system resources. The network namespaces, for example, isolate the networking resources, such as IP addresses, and IP routing tables, in order to provide each container with a separated network stack.

\begin{figure}[t]
	\begin{center}
		\scalebox{.4}{\includegraphics{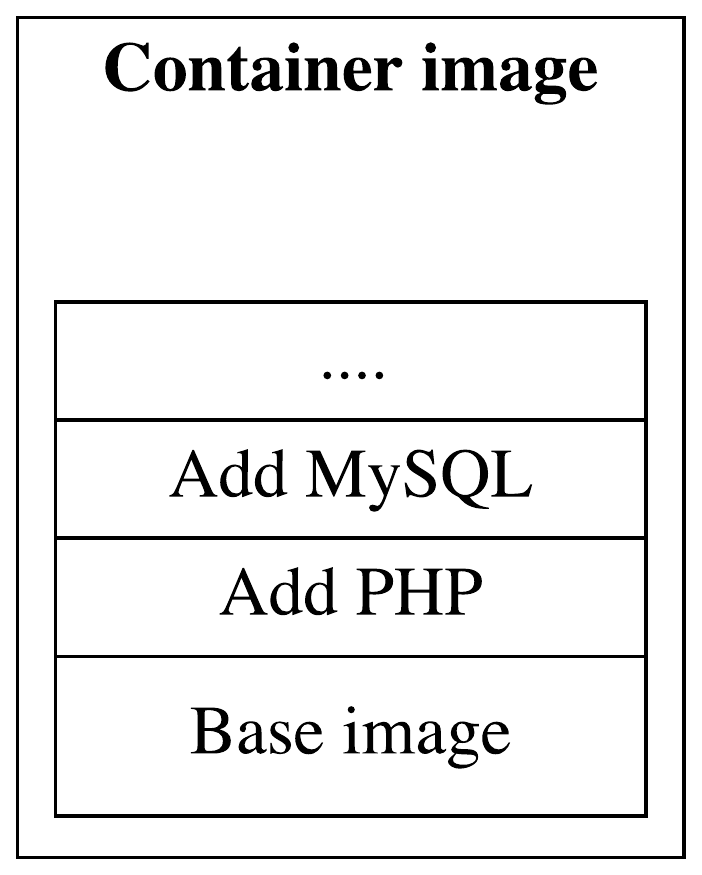}}
		\caption{Container image}
		\label{fig:container_image}
	\end{center}
\end{figure}

Docker launches its containers from \textit{Docker images}. A Docker image is a series of data layers on top of a base image (Fig.~\ref{fig:container_image}). Every Docker image starts from a base image, such as Ubuntu base image or OpenSuse base image. When users make changes to a container, instead of directly writing the changes to the image of the container, Docker adds an additional layer containing the changes to the image. For example, if the user installs MySQL to an Ubuntu image, Docker creates a data layer containing MySQL and then adds to the image. This process makes the image distribution process more efficiently since only the update needs to be distributed. 

In order to work with multiple layers of an image as it were a single file system layer, Docker uses a special file system called \textit{Union File System (UnionFS)}. It allows files and directories in different file systems to be combined into a single consistent file system.

\subsection{Docker Hub}
Docker hub \cite{_docker_????} is a central repository of images (both public and private), via which users can share their customized images. Users can also search for published images and download them with the Docker client. Furthermore, users can verify the authenticity and integrity of the downloaded images since Docker signed and verified the images when their owner submitted them to the hub.


\section{Docker Security Analysis}
\label{sec:docker_security}
Security is one of the major challenges when running services in virtual environments, especially in a multi-tenant cloud system. Virtual machines provided by hypervisor-based virtualization techniques are claimed to be more secure than containers as they add an extra layer of isolation between the applications and the host. An application running inside a VM is only able to communicate with the VM kernel, not the host kernel. Consequently, in order for the application to escalate out of a VM, it must bypass the VM kernel and the hypervisor before it can attack the host kernel. On the other hand, containers can directly communicate with the host kernel, thus allowing an attacker to save a great amount of effort when breaking into the host system. This raises a security concern over containers.

Docker is also a container-based virtualization technologies, thus having the same issue. Our analysis aims to discover whether Docker provides a safe environment to run applications. The analysis considers two areas: the internal security of Docker containers and how Docker containers interact with the additional security systems of the kernel.

\subsection{Docker Internal Security}
We examine the internal security of Docker based on the system and attacker model and security requirements as described by Reshetova et al. \cite{reshetova_security_2014} for comparing a number of the OS-level virtualization technologies. 

The system and attacker model is as follows: A single host machine is running a number of Docker containers \textit{$c_1$ ... $c_n$}, on which a subset \textit{$\overline{C}$} of the containers are compromised and the attacker has full control over those, but the remaining subset of containers \textit{$C$} is still under the control of the legitimate users. In this model, the attacker can perform various types of attacks, such as Denial-of-Service and Privilege escalation. 

In order to encounter with these attacks, the authors stated that an OS-level virtualization solution should satisfy the following requirements: process isolation, filesystem isolation, device isolation, IPC isolation, network isolation and limiting of resources. The next sections present our analysis on how Docker fulfills the requirements.

\subsubsection*{Process Isolation}
The main goal of process isolation is to prevent compromised containers from using process management interfaces to interfere with other containers. Docker achieves isolation of processes by wrapping the processes running in containers into namespaces and limiting their permissions and visibility to processes running in the other containers and the underlying host.

This mechanism operates with the support of the \textit{PID namespaces}, which isolate the process ID number space of a container from that of the host. Since PID namespaces are hierarchical \cite{_pid_????}, a process can only see the other processes in its own namespace or in its "children" namespaces. As a consequence, once a new namespace is created and assigned to a container, the host can observe and affect the processes inside the new PID namespace of the container, but the processes inside the container cannot observe or do anything to the other processes running in the host or in other containers. If the attacker cannot observe other processes, it is harder to attack them. 

The PID namespaces also allow each container to have its own init-like process (PID 1), which causes all the processes in a namespace to be terminated if it is terminated. This process assists the administrator in completely shutting down a container when something suspicious is detected.  

\subsubsection*{Filesystem Isolation}
In order to achieve filesystem isolation, the filesystems of the host and containers must be protected from illegitimate access and modification. 

Docker uses the \textit{mount namespaces}, also called the \textit{filesystem namespaces}, to isolate the filesystem hierarchy associated with different containers. The mount namespaces provide the processes of each container a different view of the filesystem tree and restrict all the mount events occurring inside the container to only have impact inside the container. However, some of the kernel filesystems are not namespaced; for example, those under $/sys$, $/proc/sys$, $/proc/sysrq-trigger$, $/proc/irq$, and $/proc/bus$, and a Docker container needs to mount them in order to operate. This causes the issue that a container inherits the view of these filesystems from the host and are able to access them directly. Docker limits the threats that a compromised container could make to the host via these filesystems with the two filesystems protection mechanisms: (1) removing the write permission to these filesystems from containers and (2) not allowing any process of a container to remount any filesystem within the container \cite{j._walsh_bringing_????}. The second mechanism is achieved by removing the $CAP\_SYS\_ADMIN$ capability from containers.

Docker also employs a mechanism called \textit{copy-on-write} file system \cite{j._walsh_bringing_????}. As mentioned before, Docker creates containers based on file system images, and a container can write content to its own base image. When multiple containers are created on the same image, the copy-on-write file system allows each container to write content to its specific file system, thus preventing other containers from discovering the changes occurring inside the container.  

\subsubsection*{Device Isolation}
In Unix, the kernel and applications access the hardware through device nodes which basically are special files acting as the interfaces to the device drivers. If a container can access some important device nodes, such as $/dev/mem$ (the physical memory), $/dev/sd*$ (the storage) or $/dev/tty$ (the terminal), it can make serious damage on the host system. Thus, it is crucial to limit the set of device nodes that a container can access. 

The \textit{Device Whitelist Controller} feature \cite{_device_????} of cgroups provides means to limit the set of devices that Docker allows a container to access. It also prevents the processes in containers from creating new device nodes. Furthermore, Docker mounts container images with \textit{nodev}, meaning that even if a device node was pre-created inside the image, the processes in the container using the image cannot use it to communicate with the kernel. By default, Docker does not give extended privileges to its containers. Therefore, they cannot access any devices. However, if the operator executes a container as "privileged", Docker grants access to all devices to the container.   
  
\subsubsection*{IPC Isolation}    
The IPC (inter-process communication) is a set of objects for exchanging data among processes, such as semaphores, message queues, and shared memory segments. The processes running in containers must be restricted so that they can communicate only via a certain set of IPC resources and are disallowed to interfere with those in other containers and the host machine. 

Docker achieves IPC isolation by using the \textit{IPC namespaces}, which allows the creation of separated IPC namespaces. The processes in an IPC namespace cannot read or write the IPC resources in other IPC namespaces. Docker assigns an IPC namespace to each container, thus preventing the processes in a container from interfering with those in other containers.

\subsubsection*{Network Isolation}
Network isolation is important to prevent network-based attacks, such as Man-in-the-Middle (MitM) and ARP spoofing. Containers must be configured in such a way that they are unable to eavesdrop on or manipulate the network traffic of the other containers nor the host. 

For each container, Docker creates an independent networking stack by using \textit{network namespaces}. Therefore, each container has its own IP addresses, IP routing tables, network devices, etc. This allows containers to interact with each other through their respective network interfaces, which is the same as how they interact with external hosts.  

\begin{figure}[t]
	\begin{center}
		\scalebox{.6}{\includegraphics{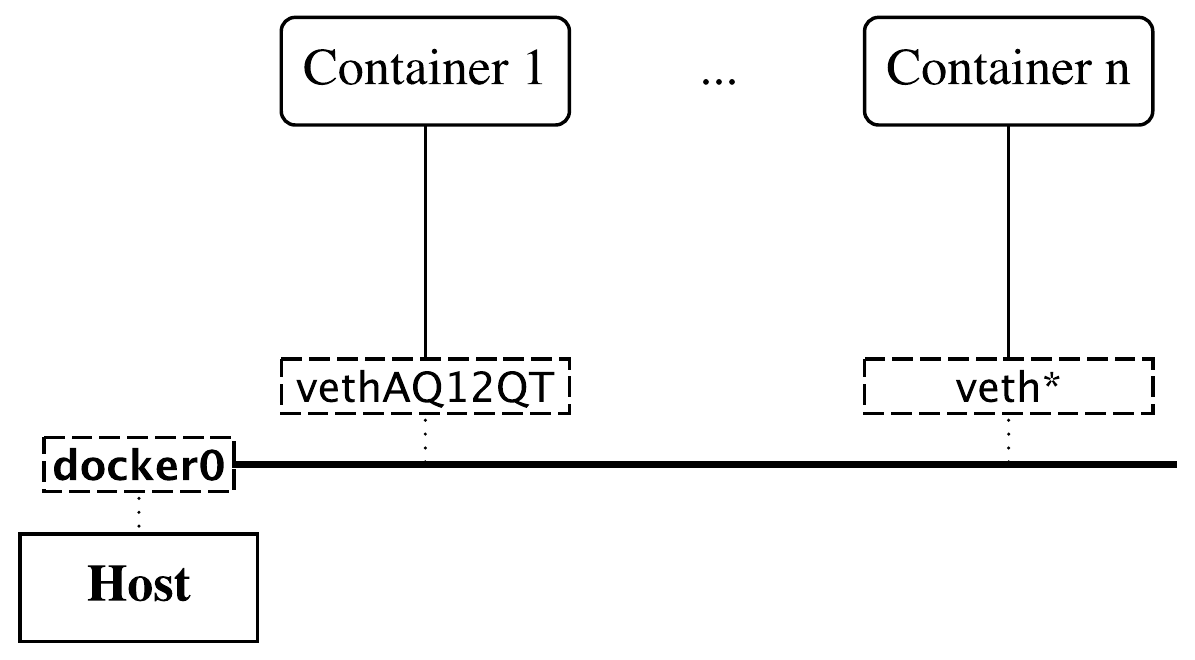}}
		\caption{The default networking model of Docker}
		\label{fig:docker_default_networking}
	\end{center}
\end{figure}

By default, connectivity between containers as well as to the host machine is provided using Virtual Ethernet bridge \cite{_docker:_????} (Fig.~\ref{fig:docker_default_networking}). With this approach, Docker creates a virtual ethernet bridge in the host machine, named $docker0$, that automatically forwards packets between its network interfaces. When Docker creates a new container, it also establishes a new virtual ethernet interface with a unique name and then connects this interface to the bridge. The interface is also connected to the $eth0$ interface of the container, thus allowing the container to send packets to the bridge. 

We note here that the default connectivity model of Docker is vulnerable to ARP spoofing and Mac flooding attacks since the bridge forwards all of its incoming packets without any filtering. 

\subsubsection*{Limiting of Resources}
Denial-of-Service (DoS) is one of the most common attacks on a multi-tenant system, where a process or a group of processes attempt to consume all of the resources of the system, thus disrupting normal operation of the other processes. In order to prevent this kind of attack, it should be possible to limit the resources that are allocated to each container.  

\textit{Cgroups} are the key component that Docker employs to deal with this issue. They control the amount of resources, such as CPU, memory, and disk I/O, that any Docker container can use, ensuring that each container obtains its fair share of the resources and preventing any container from consuming all resources. They also allow Docker to configure the limits and constraints related to the resources allocated to each container. For example, one such constraint is limiting the available CPUs available to a specific container.

\subsection{Docker and Kernel Security Systems}

Some kernel security systems exist in order to harden the security of a Linux host system, including Linux capabilities and Linux Security Module (LSM). Linux capabilities restricts the privileges assigned to each process. LSM provides a framework which allows the Linux kernel to support different security models. The LSMs that have been integrated into the official Linux kernel include AppArmor \cite{cowan_subdomain:_2000}, SELinux \cite{smalley_implementing_2001}, and Seccomp \cite{_secure_????}.

This paper surveys Linux capabilities and two LSMs, AppArmor \cite{cowan_subdomain:_2000} and SELinux \cite{smalley_implementing_2001}, which Docker currently supports. Docker can also collaborate with Seccomp but only if LXC is used; thus, we do not include it in the survey. Even though Docker does not support other security systems at the moment, it does not interfere with them. Therefore, these systems can run independently of Docker containers to protect the host \cite{_containers_????}.

\subsubsection*{Linux Capabilities}
As stated in Linux capabilities man page \cite{_linux_????}, traditionally, Unix systems classified processes into two categories: \textit{privileged processes} (owned by superuser or root) and \textit{unprivileged processes} (owned by normal users). The kernel skipped all permission checks on the privileged processes but conducted full permission checking on unprivileged processes. However, the Linux kernel, since version 2.2, divides the privileges of the superuser into \textit{capabilities}, which the kernel can independently enable or disable. 

Docker containers run on a kernel shared with the host system, so most of their tasks can be handled by the host. As a result, in most cases, it is unnecessary to provide full root privileges to a container, thus removing some of the root capabilities from a container does not affect the usability or functionality of the container but effectively improves the security of the system. For example, the $CAP\_NET\_ADMIN$ capability, which provides the ability to modify the system network, can be removed from a container since all networking configuration can be handled by the Docker daemon before starting the container. 

Docker allows configuration of the capabilities that a container can use. By default, Docker disables a large number of Linux capabilities from its containers in order to prevent an intruder to damage the host system even when the intruder has obtained root access within a container. Some of the capabilities are presented in table~\ref{tab:removed_capabilities}, and their detailed description can be found in the Linux capabilities man page \cite{_linux_????}. 

\begin{table}[t]
\begin{center}
	\scalebox{.85}{
	\begin{tabular}{ | l | l | }	
	\hline
    	$CAP\_SETPCAP$ & Modify process capabilities  \\ \hline
    	$CAP\_SYS\_MODULE$ & Insert/Remove kernel modules  \\ \hline
	$CAP\_SYS\_RAWIO$ & Modify Kernel Memory \\ \hline
	$CAP\_SYS\_PACCT$ & Configure process accounting \\ \hline
	$CAP\_SYS\_NICE$ & Modify Priority of processes \\ \hline
	$CAP\_SYS\_RESOURCE$ & Override Resource Limits \\ \hline
	$CAP\_SYS\_TIME$ & Modify the system clock \\ \hline
	$CAP\_SYS\_TTY\_CONFIG$ & Configure tty devices \\ \hline
	$CAP\_AUDIT\_WRITE$ & Write the audit log \\ \hline
	$CAP\_AUDIT\_CONTROL$ & Configure Audit Subsystem \\ \hline
	$CAP\_MAC\_OVERRIDE$ & Ignore Kernel MAC Policy \\ \hline
	$CAP\_MAC\_ADMIN$ & Configure MAC Configuration \\ \hline
	$CAP\_SYSLOG$ & Modify Kernel printk behavior \\ \hline
	$CAP\_NET\_ADMIN$ & Configure the network\\ \hline
	$CAP\_SYS\_ADMIN$ & Catch all \\ \hline
	\end{tabular}}
	\caption{Some capabilities disallowed in Docker containers \cite{j._walsh_bringing_????}}
	\label{tab:removed_capabilities}
\end{center}  
\end{table}

\subsubsection*{SELinux}
SELinux is a security enhancement to the Linux system. Linux comes with the standard \textit{Discretionary Access Controls (DAC)} mechanism (i.e., owner/group and permission flags of an object) to control the access to an object. SELinux provides an additional layer of permission checking, called \textit{Mandatory Access Control}, after the standard DAC is performed. In SELinux, everything is controlled by labels. Every file/directory, process, and system object has a label. The administrator of the system uses these labels to write rules to control access between processes and system objects. These rules are called \textit{policies}. The SELinux policies can be divided into three classes: Type enforcement, Multi-level security (MLS) enforcement, and Multi-category security (MCS) enforcement. 

With the DAC mechanism, owners have full discretion over their objects, meaning that if the owners are compromised, the attacker has control over all of their objects. In SELinux model, in contrast, the kernel manages and enforces all of the access controls over objects, not their owners. This provides a secure separation for containers as it can prevent  processes, even with root privileges, within a container to illegitimately access objects outside the containers. 

Docker uses two classes of policy enforcement: \textit{Type enforcement} and \textit{MCS enforcement}  \cite{j._walsh_bringing_????}. The Type enforcement protects the host from the processes in containers, and the MCS enforcement protects a container from another container.  

With Type enforcement, Docker labels all container processes with $svirt\_lxc\_net\_t$ type and all content within a container with $svirt\_sandbox\_file\_t$ type. The processes running with $svirt\_lxc\_net\_t$ type can only access/write to the content labeled with $svirt\_sandbox\_file\_t$ type, but not to any other label on the system. Therefore, the processes running within containers can only use the content inside containers. However, only with this policy enforcement, Docker allows the processes in one container to have access to the content of other containers. MCS enforcement is necessary to solve this issue. When a container is launched, the Docker daemon picks a random MCS label and then puts this label on all of the processes and content of the container. The kernel only allow processes to access content with the same MCS label, thus preventing a compromised process in one container from attacking other containers.   

\subsubsection*{AppArmor}
AppArmor is also a security enhancement model to Linux based on Mandatory Access Control like SELinux, but restricting its scope to individual programs. It permits the administrator to load a security profile into each program, which limits the capabilities of the program. AppArmor supports two modes: enforcement mode and complain/learning mode. The enforcement mode enforces the policies defined in the profile. However, in the complain/learning mode, the violations of profile policies are permitted, but also logged. This log can be useful for developing new profiles later.

On systems that support AppArmor, Docker provides an interface for loading a pre-defined AppArmor profile when launching a new container. This profile is loaded into the container in enforcement mode in order to ensure that the processes in the container are restricted according to the profile. If the administrator does not specify a profile when launching a container, the Docker daemon automatically loads a default profile to the container, which denies access to important filesystems on the host, such as $/sys/fs/cgroups/$ and $/sys/kernel/security/$.


\section{Discussion}
\label{sec:discuss}
The analysis shows that Docker provides a high level of isolation and resource limiting for its containers using namespaces, cgroups, and its copy-on-write file system, even with the default configuration. It also supports several kernel security features, which help to hardening the security of the host. The only problem we found with Docker was related to its default networking model. The virtual ethernet bridge which Docker uses as its default networking model, is vulnerable to ARP spoofing and MAC flooding attacks since it does not provide any filter on the network traffic passing through the bridge. However, this problem can be solved if the administrator manually adds filtering, such as ebtables \cite{_ebtables_????}, to the bridge, or changes the networking connectivity to a more secure one, such as virtual network.

It is also worth highlighting that if the operator runs a container as "privileged", Docker grants full access permissions to the container, which is nearly the same as that of processes running natively on the host. Therefore, it is more secure to operate containers as "non-privileged".

Furthermore, even though containers can provide higher density of virtual environments and better performance, they have a bigger attack surface than virtual machines since containers can directly communicate with the host kernel. However, it is possible to reduce the attack surface while maintaining these advantages. For example, this can be achieved by placing containers inside virtual machines. 


\section{Conclusion and Future work}
\label{sec:conclusion}
Container-based virtualization can provide higher density virtual environments and better performance than hypervisor-based virtualization. However, the latter is argued to be more secure than the former. In this paper, we conducted an analysis on Docker, which is one of the most popular container-based virtualization technologies, to discover how safe its containers are. Our analysis shows that Docker containers are fairly secure, even with the default configuration. The security level of Docker containers could also be increased if the operator runs them as "non-privileged" and enables additional hardening solutions in Linux kernel, such as AppArmor or SELinux.

The future work after this paper could be to compare the security of Docker containers with that of other containerization systems or with virtual machines. Such studies could lead to e.g. a detailed static analysis Docker or a broader view of security in containers in general.

\section*{Acknowledgement}
This research paper is made possible through the help and support of Miika Komu, Roberto Morabito, Jimmy Kjällman, and Tero Kauppinen from Nomadiclab. 


\bibliography{netsec_template}
\end{document}